\documentclass{aa}
\usepackage{txfonts}
\usepackage{graphicx}
\usepackage{multirow}
\usepackage{xcolor}

\def\H2{H$_2$}

\begin{document}

\title{Water and interstellar complex organics  
associated with the HH 212 protostellar disc} 
\subtitle{On disc atmospheres, disc winds, and accretion shocks}
\author{C. Codella \inst{1} 
\and 
E. Bianchi \inst{1,2,3} \and
B. Tabone \inst{4} \and 
C.-F. Lee \inst{5,6} \and
S. Cabrit \inst{4} \and
C. Ceccarelli \inst{3} \and
L. Podio \inst{1} \and
F. Bacciotti \inst{1} \and
R. Bachiller \inst{7} \and
E. Chapillon \inst{8,9} \and
F. Gueth  \inst{8} \and
A. Gusdorf \inst{4} \and
B. Lefloch \inst{3} \and
S. Leurini \inst{10} \and
G. Pineau des For$\hat {\rm e}$ts \inst{4} \and
K.L.J. Rygl \inst{11} \and
M. Tafalla \inst{7}}

\institute{
INAF, Osservatorio Astrofisico di Arcetri, Largo E. Fermi 5,
50125 Firenze, Italy
\and
Universit\`a degli Studi di Firenze, Dipartimento di Fisica e Astronomia,
Via G. Sansone 1, I-50019 Sesto Fiorentino, Italy
\and
Univ. Grenoble Alpes, Institut de
Plan\'etologie et d'Astrophysique de Grenoble (IPAG), 38401 Grenoble, France
\and
LERMA, UMR 8112 du CNRS, Observatoire de Paris, \'Ecole  Normale Sup\'erieure, 61 Av. de l'Observatoire, 75014
Paris, France
\and
Academia Sinica Institute of Astronomy and Astrophysics, P.O. Box 23-141, Taipei 106, Taiwan
\and
Graduate Institute of Astronomy and Astrophysics, National Taiwan University, No. 1, 
Sec. 4, Roosevelt Road, Taipei 10617, Taiwan
\and
IGN, Observatorio Astron\'omico Nacional, Alfonso XII 3, 28014 Madrid, Spain
\and
IRAM, 300 rue de la Piscine, 38406 Saint-Martin-d'H\`eres, France
\and
Laboratoire d'astrophysique de Bordeaux, Univ.Bordeaux,CNRS, B18N, 
all\'ee Geoffroy Saint-Hilaire, 33615 Pessac, France
\and
INAF-Osservatorio Astronomico di Cagliari, Via della Scienza 5, I-09047, Selargius (CA), Italy
\and
INAF - Istituto di Radioastronomia \& Italian ALMA Regional Centre, Via P. Gobetti 101, I-40129 Bologna, Italy
}

\offprints{C. Codella, \email{codella@arcetri.astro.it}}
\date{Received date; accepted date}

\authorrunning{Codella et al.}
\titlerunning{Water and iCOMs around HH 212 mm}

\abstract
{The unprecedented combination of high-sensitivity and high-angular resolution provided 
by the ALMA interferometer allows us to shed light  
on the processes leading to the formation
of the jet-disc system associated with a Sun-like mass protostar.}
{We investigate the physical and chemical properties of the gas associated
with water and interstellar complex organic molecules around a protostar
on solar system scales.}
{The HH 212 protostellar system, in Orion B, has been
mapped thanks to 
ALMA-Band 7 Cycle 1 and Cycle 4 observations of dueterated water (HDO) and acetaldehyde (CH$_3$CHO) 
emission with an angular resolution down to $\sim$0$\farcs$15 (60 au).}
{Many emission lines due to 14 CH$_3$CHO and 1 HDO transitions
at high excitation ($E_{\rm u}$ between 163 K and 335 K) have
been imaged in the inner $\sim$ 70 au region.
The local thermal equilibrium analysis of the CH$_3$CHO emission leads to
a temperature of 78$\pm$14 K and a column density
of 7.6$\pm$3.2 $\times$ 10$^{15}$ cm$^{-2}$, which, when $N_{\rm H_2}$ of 10$^{24}$ cm$^{-2}$ is assumed, leads to  
an abundance of 
$X_{\rm CH_3CHO}$ $\simeq$ 8 $\times$ 10$^{-9}$.
The large velocity gradient analysis of the HDO emission
also places severe constraints on the volume density,
n$_{\rm H_2}$ $\geq$ 10$^8$ cm$^{-3}$. The line profiles
are 5--7 km s$^{-1}$ wide, and
CH$_3$CHO and HDO both show a $\pm$ 2 km s$^{-1}$ velocity
gradient over a size of $\sim$ 70 au (blue-shifted emission towards  
the north-west and red-shifted emission towards the south-east)
along the disc equatorial
plane, in agreement with what was found so far using other
molecular tracers.} 
{The kinematics of CH$_3$CHO and HDO are consistent with the occurrence of 
a centrifugal barrier,
that is, the infalling envelope-rotating disc ring, which
is chemically enriched through low-velocity accretion shocks. 
The emission radius is $\sim$ 60 au, in good agreement with 
what was found before for another interstellar complex organic molecule
such as NH$_2$CHO. We support a vertical structure for the
centrifugal barrier, suggesting 
the occurrence of two outflowing, expanding,
and rotating rings above and below (of about 40-45 au) the optically thick
equatorial disc plane. It is tempting to speculate
that these rings could probe the basis of a wind launched
from this region.}


\maketitle

\section{Introduction}

Imaging of molecular emission from Class 0   
protostars is fundamental for understanding
the steps needed to reach more evolved stages,
eventually associated with protoplanetary systems
(e.g. Ceccarelli et al. 2007, 2015; Herbst \& van Dishoeck 2009; 
Tobin et al. 2012; J\o{}rgensen 2016, and references therein).
The recent advent of ALMA is making a great contribution, among other topics, 
to the study of the
pristine jet-disc systems associated 
with Sun-like protostars (e.g. Murillo et al. 2013; Harsono et al. 2015).
More specifically,  
the unprecedented combination of sensitivity and high-spatial
resolution finally allows us to distinguish the numerous kinematical
components at work in the inner 50 au,
which is the region where a planetary system is expected to form.
In this context, astrochemical studies are crucial for imaging the transition
from an infalling-rotating envelope to a rotating disc, which
occurs around protostars on spatial scales of 30 to 100 au. 
Slow (less than 2 km s$^{-1}$) accretion shocks 
are expected at the envelope-disc surface, and a consequent
drastic enrichment of the gas-phase material is the natural consequence
(Sakai et al. 2014ab, 2016, 2017; Oya et al. 2016; Lee et al. 2016, 2017b). 
Among other species, this effect concerns  
the so-called interstellar complex organic molecules
(iCOMs; C-bearing species with more than six atoms) that can be considered key bricks to the next step of complex prebiotic chemistry, and it is also relevant for water, another 
obvious key species for this type of studies.
Despite the high spatial resolution observations 
obtained in this context in the past
years, 
there are still hotly debated questions. 
Do the iCOMs associated with the disc surface layers 
leave the disc itself (see Lee et al. 2017b)? 
If so, are iCOMs
associated with wide-angle slow disc winds?  
Further high-angular resolution observations 
of Class 0 environments are needed to
advance in this type of studies.

HH 212 can be considered one of the best interstellar laboratories 
in which to study in detail the chemistry and physics of 
pristine jet-disc systems that are associated
with a forming star.
The HH 212 mm Class 0 protostar 
($L_{\rm bol}$ $\sim$ 9 $L_{\rm \odot}$, $M$ $\simeq$ 0.2--0.3
$M_{\rm \odot}$) 
is located in the L1630 cloud in Orion, between
NGC 2068 and NGC 2024, at a distance that has recently been estimated to
be 405$\pm$15 pc (Kounkel et al. 2017; Kounkel, private communication). 
HH 212 is associated with an extended symmetric molecular and rotating jet 
discovered using H$_2$ by Zinnecker et al. (1998), which was later extensively 
observed using interferometers such as IRAM-NOEMA, the SMA, and
ALMA
(e.g. Lee et al. 2006, 2007, 2008, 2014, 2015, 2017c; Codella et al. 2007;
Cabrit et al. 2007, 2012; Podio et al. 2015).
A dusty disc with a radius of $\sim$ 60 au  
that is deeply embedded in 
an infalling-rotating
flattened envelope has recently been imaged with 
an unprecedented spatial resolution of 8 au using ALMA Band 7 continuum
emission by Lee et al. (2017a).
Using molecular (HCO$^+$, C$^{17}$O, SO) emission, the
disc kinematics has also been investigated and revealed a velocity 
gradient that agrees with the jet rotation on scales $\leq$ 100 AU 
(Lee et al. 2014; Codella et al. 2014; Podio et al. 2015) and  
on scales of 40--50 au (Lee et al. 2017b).   
Finally, based on emission from CH$_3$OH and S-bearing species, 
   Codella et al. (2014), Leurini et al. (2016), and
Bianchi et al. (2017) speculated that there might be a disc wind. This was later discussed in detail by
Tabone et al. (2017) and Lee et al. (2017d). 

The chemical enrichment of the HH 212 mm surroundings has  
been investigated by Codella et al. (2016) and Lee et al. (2017b), 
who reported deuterated
water (HDO), acetaldehyde (CH$_3$CHO), and formamide (NH$_2$CHO) 
in the inner 40-50 au 
from the protostar. This revealed the first hot corino in Orion.
In particular, Lee et al. (2017b) showed that iCOMs 
trace the surface layers of the rotating disc; these layers are
also called the disc atmosphere. 
Here we further investigate the origin
of the chemical complexity in the HH 212 mm hot corino by 
imaging the spatial distribution of water (in its deuterated form) 
and CH$_3$CHO 
down to an angular scale of 150 mas.

\begin{figure*}
\centerline{\includegraphics[angle=0,width=18cm]{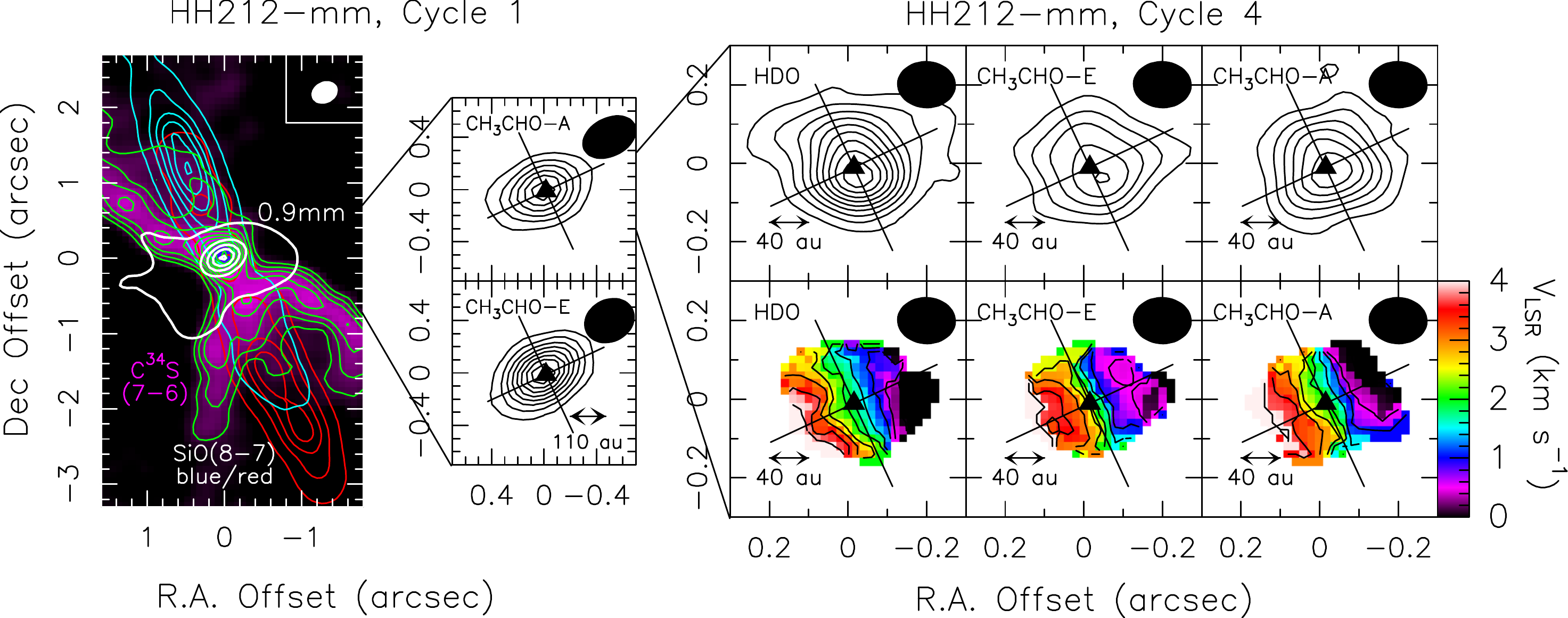}}
\caption{{\it Left panel:}  HH 212 protostellar system
as observed by ALMA-Band 7 during Cycle 1.
Blue and red contours plot the blue- and red-shifted SiO(8--7) jet
and the C$^{34}$S(7--6) asymmetric cavity (magenta
and green contours)
overlaid on the continuum at 0.9 mm (white contours).
Positions are given with respect to the coordinates
of the MM1 protostar, reported in Sect. 2.
The filled ellipse shows
the synthesised beam (HPBW) for the continuum:
$0\farcs36\times0\farcs28$ (--62$\degr$).
The beam for the SiO and C$^34$S images is
$0\farcs43\times0\farcs33$ (--64$\degr$).
First contours and steps for the continuum are are
5$\sigma$ (1.5 mJy beam $^{-1}$) and 60$\sigma$, respectively.
For SiO, the first contours and steps are 5$\sigma$
and 25$\sigma$, respectively: the blue map
has been obtained by integrating down to
--21 km s$^{-1}$ (1$\sigma$ = 29 mJy beam $^{-1}$ km s$^{-1}$),
and the red map
collects emission up to +12 km s$^{-1}$ (1$\sigma$ = 23
mJy beam $^{-1}$ km s$^{-1}$).
The C$^{34}$S map has been obtained by integrating
the velocities from -1 km s$^{-1}$ to +1 km s$^{-1}$ with respect to
the systemic velocity $v_{sys}$ = +1.7 km s$^{-1}$ (Lee et al. 2014):
the first contour and steps are
5$\sigma$ (10 mJy beam $^{-1}$ km s$^{-1}$) and 3$\sigma$, respectively.
{\it Central panels:} Zoom-in of the central region
as observed by ALMA Band 7 Cycle 1:
CH$_3$CHO(18$_{1,17}$--17$_{1,16}$)E and
CH$_3$CHO(18$_{4,15}$--17$_{4,14}$)A emission integrated
over $\pm$5 km s$^{-1}$ with respect to $v_{sys}$
(black contours).
The first contours and steps are 3$\sigma$
(18 mJy beam $^{-1}$ km s$^{-1}$).
The HPBWs are
$0\farcs41\times0\farcs33$ (--64$\degr$) for
CH$_3$CHO(18$_{1,17}$--17$_{1,16}$)E,
and $0\farcs44\times0\farcs33$ (--63$\degr$)
for CH$_3$CHO(18$_{4,15}$--17$_{4,14}$)A.
The black triangle shows the MM1 coordinates, and
the tilted cross indicates the directions of the jet
and of the equatorial plane.
{\it Right panels:}
Further zoom-in of the central region, as observed by ALMA Band 7 Cycle 4,
showing the HDO(3$_{\rm 3,1}$--4$_{2,2}$),
CH$_3$CHO(18$_{\rm 0,18}$--17$_{\rm 0,17}$)E,
and CH$_3$CHO(18$_{\rm 0,18}$--17$_{\rm 0,17}$)A.
The emission was integrated over 10 km s$^{-1}$ around $v_{sys}$
(upper panels; black contours).
The first contours and steps are 3$\sigma$
(14 mJy beam $^{-1}$ km s$^{-1}$).
The HPBW is $0\farcs15\times0\farcs12$ (PA = -88$\degr$).
The corresponding first-moment maps are reported in colour scale in the
lower panels. Contours are from 0 km s$^{-1}$ to +4 km s$^{-1}$
in steps of 0.5 km s$^{-1}$.}
\label{allmaps}
\end{figure*}

\section{Observations} 

The HH 212 protostellar system was observed 
during ALMA Cycle 1 in Band 7 using 34 12m antennas  
between 15 June and 19 July 2014.
In addition, HH 212 was also observed in Band 7 
during Cycle 4 using 44 12m antennas
between 6 October and 26 November 2016.
The maximum baselines were 650 m for Cycle 1 and 3 km 
for Cycle 4, while the maximum unfiltered scale is $\sim$ 3$\arcsec$, that is,
1200 au.  

\begin{figure*}
\centerline{\includegraphics[angle=0,width=14cm]{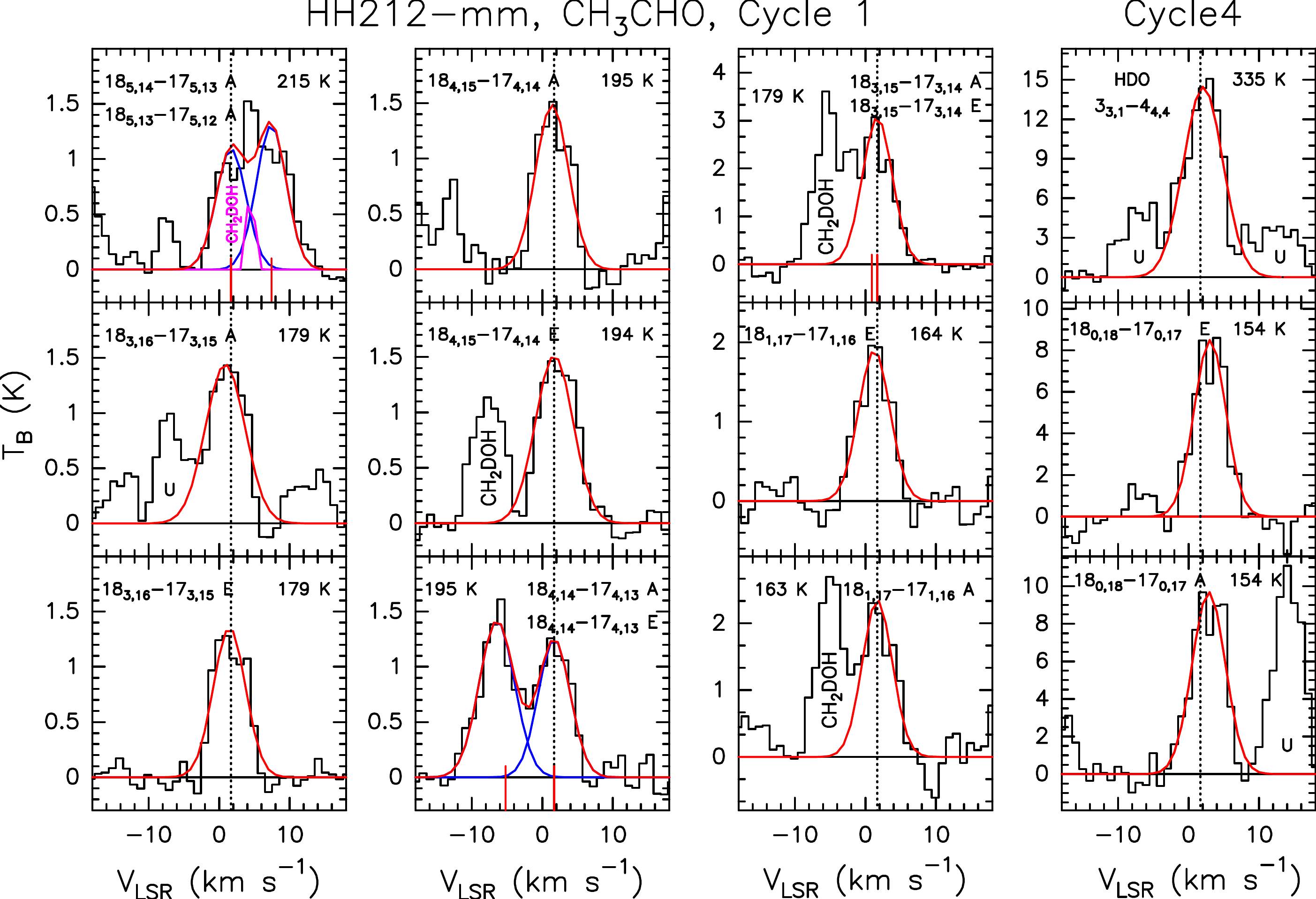}}
\caption{CH$_3$CHO and HDO line profiles in $T_{\rm b}$ scale 
(with a spectral resolution of 1 km s$^{-1}$) observed during
ALMA Cycle 1 and 4 (right panels) operations. Species, transitions, and upper level excitations ($E_{\rm u}$ in K)
are reported (see Table 1). The vertical
dashed line stands for the systemic velocity $v_{\rm sys}$ = +1.7 km s$^{-1}$ (e.g. Lee et al. 2014).
Red curves are Gaussian fits, the results of which are reported in Table 1.
In some of the panels, vertical red segments indicate an additional
CH$_3$CHO line close to that centred at $v_{\rm sys}$. 
The CH$_2$DOH labels (see also the magenta profile in the
upper left panel, residual of the
(18$_{\rm 5,14}$--17$_{\rm 5,13}$)A and (18$_{\rm 5,13}$--17$_{\rm 5,12}$)A
fit) are for 
emission lines of deuterated methanol, 
published by Bianchi et al. (2017). Some unidentified
lines (U labels) are also reported.
}
\label{spectra}
\end{figure*}

In Cycle 1, two spectral windows (337.1--338.9 GHz and 348.4--350.7 GHz) 
were simultaneously observed using spectral channels 
of 488 kHz (0.42--0.43 km s$^{-1}$), 
and then smoothed to 1.0 km s$^{-1}$ to increase sensitivity
and obtain a higher signal-to-noise ratio (S/N) for the CH$_3$CHO lines.
The spectral windows were also used to derive continuum from the
line-free channels.
Standard procedures were used to calibrate the data,  
using quasars J0607-0834, J0541--0541, J0423--013, and Ganymede.
The continuum-subtracted images have
clean-beam FWHMs around $0\farcs41\times0\farcs33$ to
$0\farcs44\times0\farcs34$
(PA = --63$\degr$) and an r.m.s. noise level of
$\sim$ 2 mJy beam$^{-1}$ in the 1 km s$^{-1}$ channels.  
During Cycle 4, we used a single spectral  
window between 334.1--336.0 GHz with a 0.42 km s$^{-1}$ spectral
resolution, successively smoothed to 
1 km s$^{-1}$ (this was also used to collect continuum emission)
to improve the sensitivity.
Calibration was carried out
using quasars J0510+1800, J0552+0313, J0541--0211, and J0552--3627
\footnote{Errors on absolute positions of these sources at these 
frequencies are smaller than 1 mas (https://almascience.eso.org/sc/).}.
The continuum-subtracted images have a typical 
clean-beam FWHM of $0\farcs15\times0\farcs12$
(PA = -88$\degr$) and an r.m.s. noise level of
$\sim$ 1 mJy beam$^{-1}$ per channel.   

Spectral line imaging was achieved with the 
CASA\footnote{http://casa.nrao.edu}  package, while 
data analysis was performed using the 
GILDAS\footnote{http://www.iram.fr/IRAMFR/GILDAS} package. 
Positions are given with respect to the MM1 protostar continuum peak
located at $\alpha({\rm J2000})$ = 05$^h$ 43$^m$ 51$\fs$41, 
$\delta({\rm J2000})$ = --01$\degr$ 02$\arcmin$ 53$\farcs$17
(e.g. Lee et al. 2014).
HDO and CH$_3$CHO emission lines (see Table 1) were identified using 
spectroscopic parameters extracted from the Jet Propulsor Laboratory  
(JPL\footnote{https://spec.jpl.nasa.gov/}, Pickett et al. 1998)
molecular database. We also used the Cycle 1 SiO(8--7) and C$^{34}$S(7--6) 
emission at 347330.63 MHz
and 337396.69 MHz (also from JPL), respectively, as well as the continuum 
to show the jet direction, the asymmetric cavity, 
and the surrounding envelope  
(Fig. 1; see also Tabone et al. 2017).

\section{Spectra and images}

Our ALMA dataset allows us to detect 14 CH$_3$CHO
emission lines (12 in Cycle 1, and 2 in Cycle 4). Figure 1 shows the spectra 
extracted from the protostellar
position: the profiles, fit using the GILDAS
package (see Table 1; note that in two cases, 2 CH$_3$CHO lines
are blended), are Gaussian-like and the S/N 
of the $T_{\rm peak}$ is always higher than 6. 
The line excitation is high, with $E_{\rm u}$ from 163 K to 215 K. 
In addition, we detect in the Cycle 4 spectra the 3$_{3,1}$--4$_{2,2}$ line
of deuterated water (HDO), with $E_{\rm u}$ = 335 K (S/N $\sim$ 12).
All the emission lines peak at velocities close to 
the systemic velocity $v_{\rm sys}$ = +1.7 km s$^{-1}$ (Lee et al. 2014) 
and their FWHM line widths 
are between +5.1 km s$^{-1}$ and +6.6 km s$^{-1}$ , in agreement with 
the same CH$_3$CHO and HDO lines
detected at lower angular resolution by Codella et al. (2016).
We note that the fluxes as integrated over the whole velocity range are
measured with a dynamical range of 10--50.

Figure 1 (middle panels) reports 
the spatial distribution of the  
CH$_3$CHO(18$_{1,17}$--17$_{1,16}$)E and
CH$_3$CHO(18$_{4,15}$--17$_{4,14}$)A emission 
(observed in Cycle 1) integrated
over $\pm$5 km s$^{-1}$ with respect to $v_{sys}$.
These maps are representative of the spatial
distributions of all the CH$_3$CHO lines detected in Cycle 1.
The CH$_3$CHO emission was not spatially resolved when observed
with a beam of $\sim$ 0$\farcs$6 by Codella et al. (2016), 
and our Cycle 1 maps ($\sim$ 0$\farcs$4) are still  
spatially unresolved. 
The Cycle 4 images with higher spatial resolution  
(see Fig. 1) allow us to resolve the size of the total emitting regions,
obtained with 2D Gaussian fits (after deconvolving the beam size):
161$\pm$17 mas (HDO), 205$\pm$25 mas (CH$_3$CHO-A), and 213$\pm$34 mas (CH$_3$CHO-E).
The average is then $\sim$ 0$\farcs$18, that is, 73 au.

\begin{figure}
\centerline{\includegraphics[angle=0,width=8.7cm]{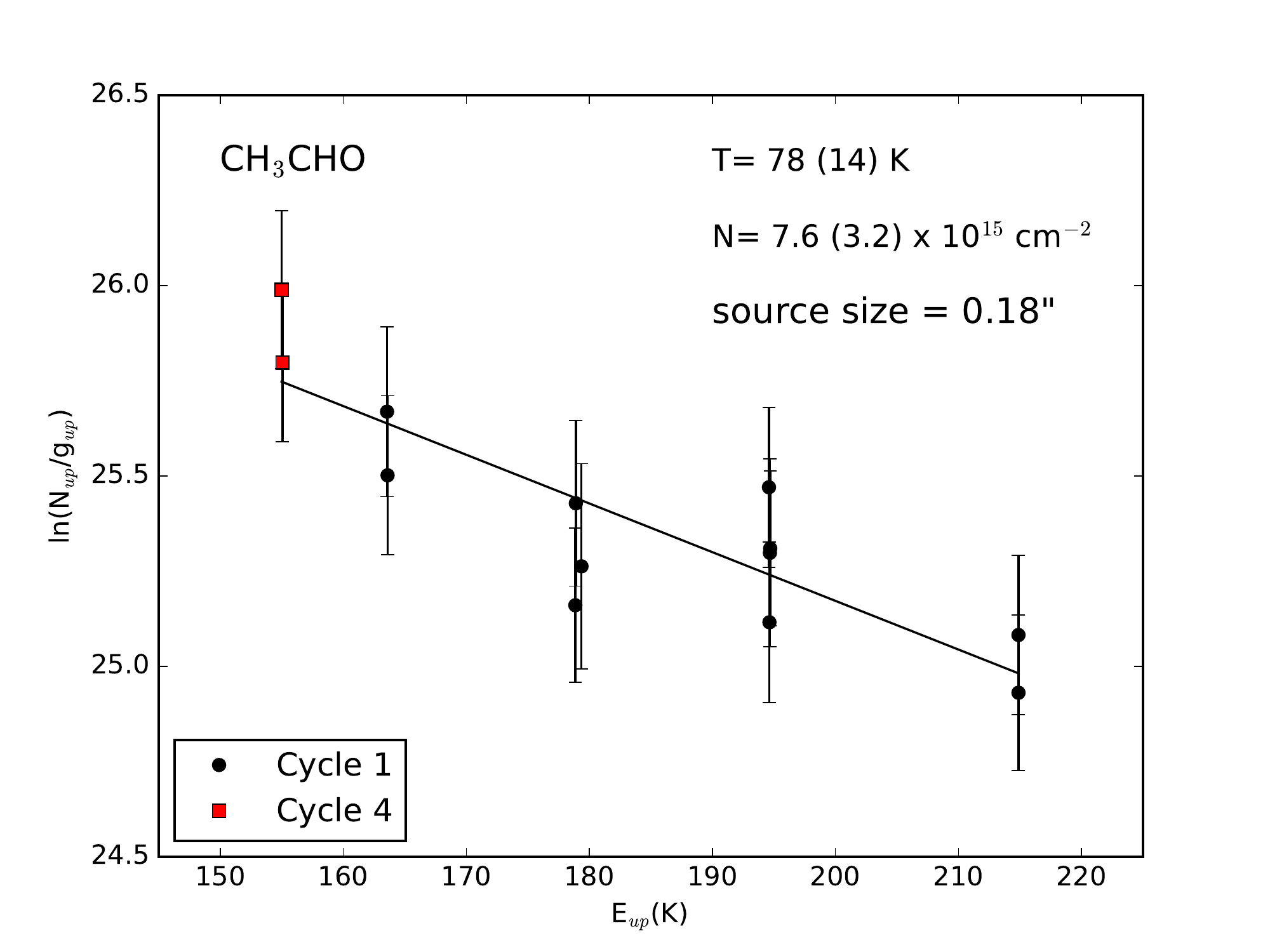}}
\centerline{\includegraphics[angle=0,width=8cm]{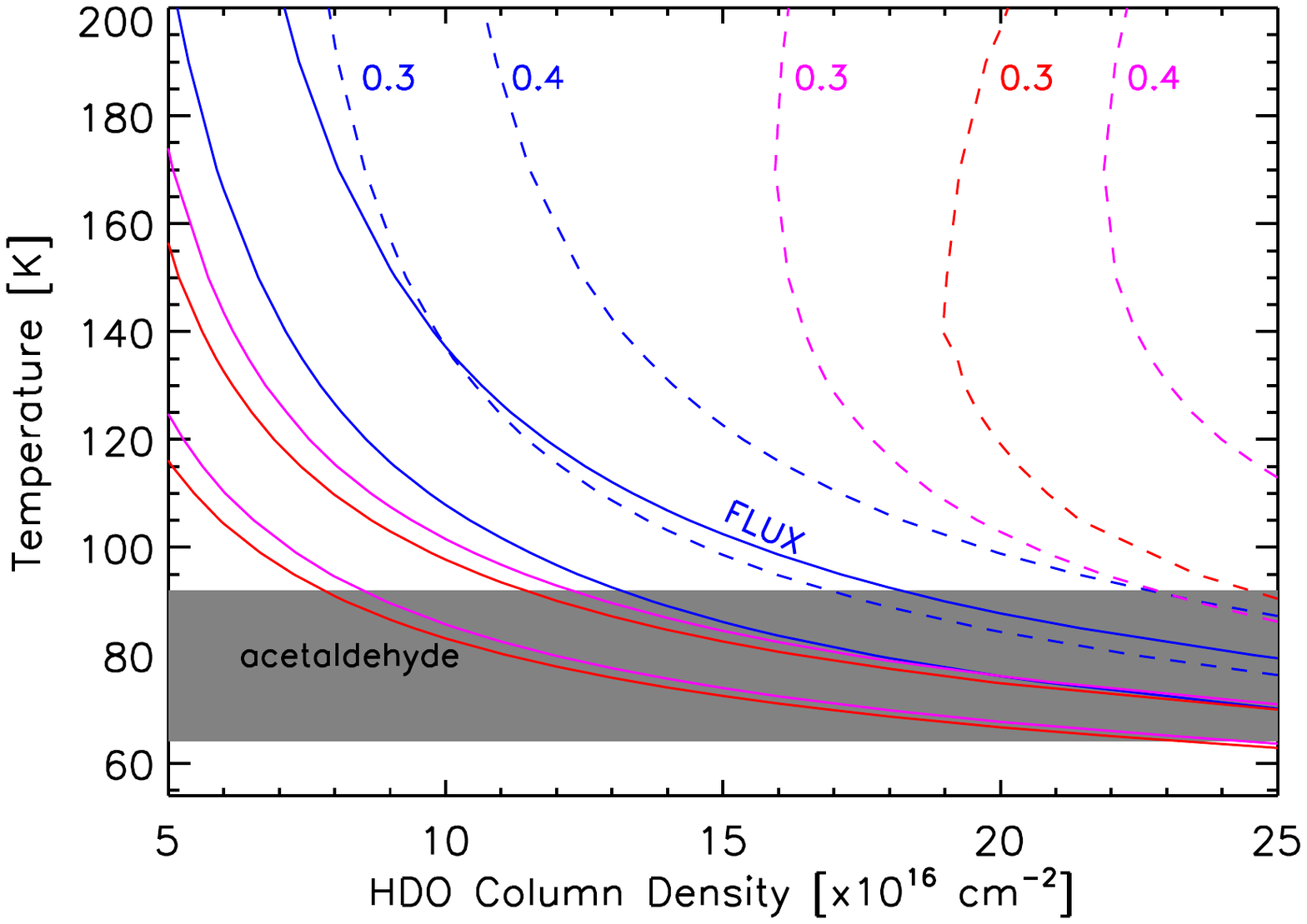}}
\caption{{\it Upper panel:} Rotation diagrams for CH$_{3}$CHO, derived using Cycle 1 (black points) and Cycle 4 (red) data.
A total emitting region size of $0\farcs18$ is assumed (see text).
The parameters $N_{\rm u}$, $g_{\rm u}$, and $E_{\rm up}$ are the column density, degeneracy, and energy
of the upper level, respectively. The derived values of the rotational
temperature and total column density are reported.
{\it Lower panel:} LVG predictions of the temperature vs. HDO column
density required to
reproduce the observed velocity-integrated emission
(the portion of the plot delimited by solid curves; see Table 1)
densities of 10$^{8}$ cm$^{-3}$ (blue),
10$^{9}$ cm$^{-3}$ (magenta), and
10$^{10}$ cm$^{-3}$ (red). Dashed lines are for the optical depth
($\tau$ = 0.3 and 0.4), increasing with column density.
A source size of 0$\farcs$18 is assumed. The grey zone shows the
temperature derived from CH$_3$CHO (see upper panel).}
\label{rd}
\end{figure}

\begin{table*}
\caption{CH$_{3}$CHO and HDO emission lines detected towards HH 212 mm during
Cycle 1 and Cycle 4 observations.
}
\begin{tabular}{lrrcrrccc}
\hline
\multicolumn{1}{c}{Transition$^a$} &
\multicolumn{1}{c}{$\nu_{\rm 0}$ $^a$} &
\multicolumn{1}{c}{$E_{\rm u}$ $^a$} &
\multicolumn{1}{c}{$S_{ij} \mu^{2}$ $^a$} &
\multicolumn{1}{c}{rms $^b$} &
\multicolumn{1}{c}{$T_{\rm peak}$ $^b$} &
\multicolumn{1}{c}{$V_{\rm peak}$ $^b$} &
\multicolumn{1}{c}{FWHM $^b$} &
\multicolumn{1}{c}{$I_{int}$ $^b$} \\
\multicolumn{1}{c}{} &
\multicolumn{1}{c}{(GHz)} &
\multicolumn{1}{c}{(K)} &
\multicolumn{1}{c}{($D^{2} $)} &
\multicolumn{1}{c}{(K)} &
\multicolumn{1}{c}{(K)} &
\multicolumn{1}{c}{(km s$^{-1}$)} &
\multicolumn{1}{c}{(km s$^{-1}$)}&
\multicolumn{1}{c}{(K km s$^{-1}$)}  \\
\hline
\multicolumn{9}{c}{Cycle 1} \\
\hline
CH$_{\rm 3}$CHO 18$_{\rm 5,14}$--17$_{\rm 5,13}$ A & 347.2883 &          215 &  210 &  0.2& 1.3 (0.2)   & +1.7  (--)$^c$& 5.0 (--)$^c$& 6.8 (0.4) \\
CH$_{\rm 3}$CHO 18$_{\rm 5,13}$--17$_{\rm 5,12}$ A &   347.2949   & 215 &  210 &  0.1 & 1.1 (0.2)  &  +1.7 (-)$^c$& 5.0  (--)$^c$ & 5.8 (0.2) \\
CH$_{\rm 3}$CHO 18$_{\rm 3,16}$--17$_{\rm 3,15}$ A & 347.5192 & 179 & 221 & 0.2 &1.4 (0.2) & +0.9 (0.3) & 6.6 (0.7) & 10.1 (0.9) \\
CH$_{\rm 3}$CHO 18$_{\rm 3,16}$--17$_{\rm 3,15}$ E  &   347.5633 & 179 & 221 &   0.1 &1.4 (0.2) & +1.5 (0.1) & 5.1 (0.2) & 7.7 (0.2) \\
CH$_{\rm 3}$CHO 18$_{\rm 4,15}$--17$_{\rm 4,14}$ A & 347.6504 & 195 & 216 &0.4 & 1.5 (0.2) & +1.4 (0.4) & 5.5 (0.8)& 8.7 (1.3) \\
CH$_{\rm 3}$CHO 18$_{\rm 4,15}$--17$_{\rm 4,14}$ E  & 347.7563 & 194 & 213 &0.2  & 1.5 (0.2) & +1.7  (--)$^c$&  6.3 (0.5)&  10.2 (0.7) \\
CH$_{\rm 3}$CHO 18$_{\rm 4,14}$--17$_{\rm 4,13}$ E & 347.8310 & 195 &  213 & 0.1  & 1.3 (0.1) & +1.7 (0.2) & 5.3 (0.4) & 7.1 (0.5)  \\
\vspace{0.1 cm}
CH$_{\rm 3}$CHO 18$_{\rm 4,14}$--17$_{\rm 4,13}$ A & 347.8390 & 195 & 213 &  0.1 & 1.4 (0.1) & +0.5 (0.1) & 5.9 (0.3) & 8.8 (0.3)  \\
CH$_{\rm 3}$CHO 18$_{\rm 3,15}$--17$_{\rm 3,14}$ A & 350.1334 & 179 & 221  &\multirow{2}{*}{ 0.1}& \multirow{2}{*}{ 3.1 (0.1)} & \multirow{2}{*}{+1.7  (--)$^c$} & \multirow{2}{*}{5.0  (-)$^c$ } & \multirow{2}{*}{ 16.3      (2.3)  }\\
\vspace{0.1 cm}
CH$_{\rm 3}$CHO 18$_{\rm 3,15}$--17$_{\rm 3,14}$ E & 350.1344 &179 & 221 & &  &  &  &  \\
CH$_{\rm 3}$CHO 18$_{\rm 1,17}$--17$_{\rm 1,16}$ E & 350.3628 & 164 & 226 &0.2 & 1.9 (0.2) &  +1.3 (0.2) &  5.2 (0.3) & 10.6 (0.6) \\
CH$_{\rm 3}$CHO 18$_{\rm 1,17}$--17$_{\rm 1,16}$ A &    350.4458 &  163 & 226 & 0.5  & 2.4 (0.4) &  +1.7 (--)$^c$  & 5.0 (--)$^c$ & 12.5 (1.2) \\
\hline
\multicolumn{9}{c}{Cycle 4} \\
\hline
HDO 3$_{\rm 3,1}$--4$_{2,2}$ & 335.3955 & 335 & 0.4 & 0.5 & 16.2 (1.3) & +2.1 (0.1) & 6.6 (0.2) & 101.6 (1.9) \\
CH$_{\rm 3}$CHO 18$_{\rm 0,18}$--17$_{\rm 0,17}$ E & 335.3181 & 155 & 227 & 0.6 & 8.5 (1.3) & +3.0 (0.2) & 5.3 (0.2) & 47.8 (2.6) \\
CH$_{\rm 3}$CHO 18$_{\rm 0,18}$--17$_{\rm 0,17}$ A & 335.3587 & 155 & 227 & 1.1 & 9.6 (1.3) & +2.9 (0.2) & 5.6 (0.5) & 57.3 (4.5) \\
\hline
\end{tabular}

$^a$ Frequencies and spectroscopic parameters have been extracted from the Jet Propulsion Laboratory molecular database (Pickett et al. 1998) for all the transitions. Upper level energies refer to the ground state of each symmetry. $^b$ Gaussian fit. Spectral resolution of 1 km s$^{-1}$.
$^c$ Assumed. \\
\end{table*}

\section{Physical properties}

Given the lack of the collisional rate coefficients available in literature, 
the excitation temperature and column density 
of CH$_3$CHO were derived using the rotational diagram
approach (Fig. 3), 
assuming local thermodynamic equilibrium (LTE) conditions and optically thin emission.
We used Cycle 1 and Cycle 4 emission lines, 
and following the maps of Fig. 1, corrected the observed intensities
for a source size (of the total emitting region) of 0$\farcs$18.
We derived an excitation temperature of 78$\pm$14 K and column density of 7.6$\pm$3.2 $\times$ 10$^{15}$
cm$^{-2}$. The temperature agrees reasonably well with the Cycle 0 results reported
by Codella et al. (2016), and it is also consistent with the temperatures of 165$\pm$85 K and
171$\pm$52 K derived from methanol observations (covering a similar excitation range)
by Lee et al. (2017b) and Bianchi et al. (2017). 
On the other hand, the column density is higher by a factor 4 than 
what was reported by Codella et al. (2016).
This discrepancy, however, can be explained  
given the size of 0$\farcs$3 assumed in Codella et al. (2016).   
The lower column density derived using the Cycle 0 dataset
is an average over a larger assumed emission area. 
The Cycle 4 data analysed in this paper allow us to measure the
size of the emitting region, hence to obtain a 
reliable measure of the CH$_3$CHO column density and an improved
estimate of the abundance. As stressed in 
Codella et al. (2016), the continuum
emission, being optically thick (see e.g. Lee et al. 2017a, 
and references therein),
cannot be used to derive an $N$(H$_2$) estimate.
However, assuming H$_2$ column densities 
around 10$^{24}$ cm$^{-2}$, which is  
a typical value for hot corinos in Perseus and in B335 
(Taquet et al. 2015, Imai et al. 2016), we can infer
$X_{\rm CH_3CHO}$ $\simeq$ 8 $\times$ 10$^{-9}$. This is in 
good agreement with the CH$_3$CHO abundance found in a region within
a few 10 au around the B335 protostar using ALMA (2 $\times$ 10$^{-9}$;
Imai et al. 2016). 

The HDO emission was analysed using a non-LTE Large Velocity
Gradient (LVG) model
(see Ceccarelli et al. 2003), assuming collisional coefficients for the 
system HDO-H$_2$ computed by Faure et al. (2012).
A Boltzmann distribution for the ortho-to-para H$_2$ ratio,
in agreement with the Faure et al. (2012) computations, was found, 
showing that 
para-H$_2$ and ortho-H$_2$ collisional coefficients are different only
at temperatures $\ll$ 45 K.
Different to Codella et al. (2016), we can now use the measured size
of 0$\farcs$18. Figure 3 shows the LVG predictions in
the $T_{\rm kin}$--$N$(HDO) plane for densities of 
10$^8$ cm$^{-3}$ (blue), 10$^9$ cm$^{-3}$ (magenta), and 
10$^{10}$ cm$^{-3}$ (red; representing the LTE regime).
When we assume $N$(HDO)/$N$(H$_2$O) $\leq$ 0.1, 
$X_{\rm H_2O}$ = 3 $\times$ 10$^{-5}$; see Taquet et al. 2015)
and a source size of 0$\farcs$18, then Fig. 3 shows
that only densities of at least 10$^8$ cm$^{-3}$ are possible,
leading to $N$(HDO) $\leq$ 3 $\times$ 10$^{17}$ cm$^{-2}$. 
Densities of 10$^7$ cm$^{-3}$ would imply
$N$(HDO) $\leq$ 3 $\times$ 10$^{16}$ cm$^{-2}$, which are values that were ruled
out by the LVG solutions in Fig. 3.
With a density of 10$^8$ cm$^{-3}$ , the opacity of the
HDO line is 0.3, while lower values are expected for
higher densities.
The temperatures derived from CH$_3$CHO (grey region
in Fig. 3)
are consistent with the solutions found with HDO. In conclusion, our dataset allow us to further constrain
the physical properties associated 
with HDO and CH$_3$CHO:
$T_{\rm kin}$ = 78$\pm$14 K and n$_{\rm H_2}$ $\geq$ 10$^8$ cm$^{-3}$.

\begin{figure*}
\centerline{\includegraphics[angle=0,width=16cm]{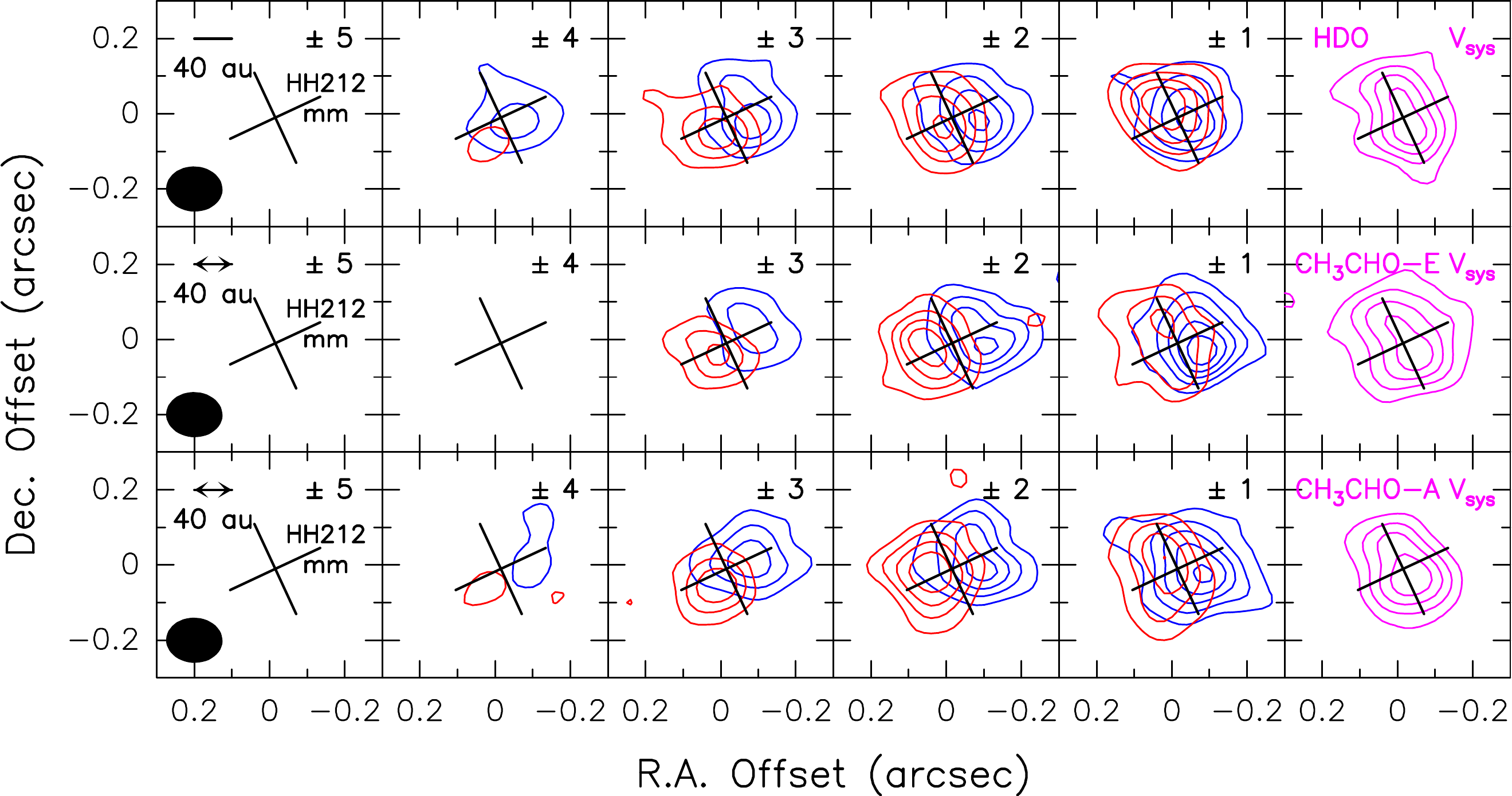}}
\caption{Channel maps of the HDO(3$_{\rm 3,1}$--4$_{2,2}$)
(upper panels) and CH$_{\rm 3}$CHO(18$_{\rm 0,18}$--17$_{\rm 0,17}$) E\&A
(lower panels)
blue- and redshifted
(continuum-subtracted) emissions observed during ALMA Cycle 4
towards the HH 212 mm protostar.
Each panel shows the emission integrated over a velocity interval
of 1 km s$^{-1}$ shifted with respect to
the systemic velocity (see the magenta channel, sampling the velocity
between +1.5 km s$^{-1}$ and +2.5 km s$^{-1}$) by the
value given in the upper right corner.
The black cross (oriented to illustrate the direction of the SiO jet and
consequently the equatorial plane, see Fig. 1)
indicates the position of the protostar.
The ellipse in all the leftmost panels shows
the ALMA synthesised beam (HPBW): $0\farcs15\times0\farcs12$ (PA = --88$\degr$).
First contours and steps correspond to 3$\sigma$
(4.5 mJy beam $^{-1}$ km s$^{-1}$ for HDO and
3.0 mJy beam $^{-1}$ km s$^{-1}$ for CH$_{\rm 3}$CHO).}
\label{channels}
\end{figure*}

\begin{figure}
\centerline{\includegraphics[angle=0,width=7cm]{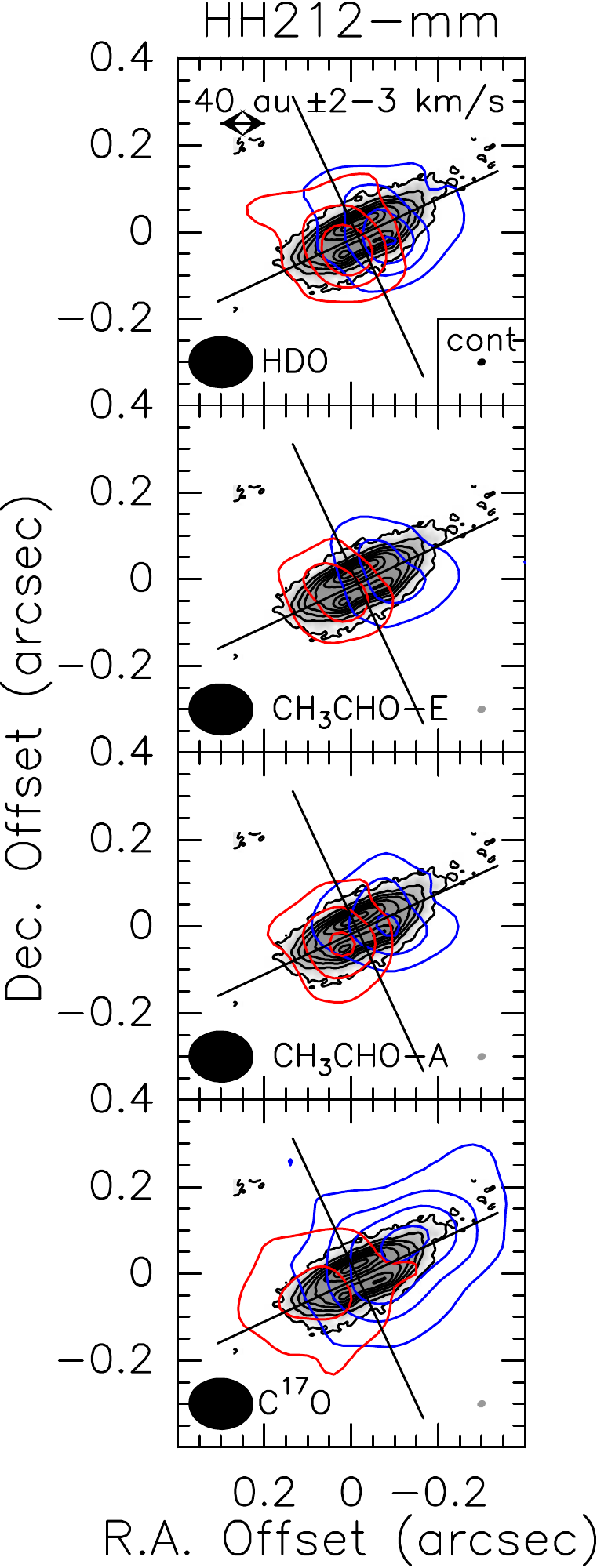}}
\caption{HDO(3$_{\rm 3,1}$--4$_{2,2}$) ({\it Upper}),
CH$_{\rm 3}$CHO(18$_{\rm 0,18}$--17$_{\rm 0,17}$) E\&A ({\it middle}), and
C$^{17}$O(3--2) ({\it lower}) channel maps
emitting $v_{\rm sys}$ $\pm$ 2--3 km s$^{-1}$ (blue and red) overlaid
on top of the disc traced by Lee et al. (2017a) using
ALMA Band 7 continuum observations (grey scale).
C$^{17}$O(3--2) emission (a tracer of the HH 212 disc at these velocities,
see also Codella et al. 2014) has been also observed in
our Cycle 4 dataset (Tabone et al. 2017).
The first contours and steps correspond to 3$\sigma$
(60 mJy beam$^{-1}$ km s$^{-1}$ for HDO and CH$_{\rm 3}$CHO, and 73
mJy beam$^{-1}$ km s$^{-1}$ for C$^{17}$O) and 5$\sigma$,
respectively.}
\label{super}
\end{figure}

\section{Kinematics}

The study of kinematics is instructive:
Fig. 1 (lower panels) reports the distribution of the first moment
of the HDO and CH$_3$CHO Cycle 4 images.  
We detect a velocity gradient ($\pm$ 2 km s$^{-1}$ with respect 
to the systemic velocity) directed along the
equatorial plane over a $\sim$ 70 au scale, 
consistent with the gradients detected 
in the envelope and in the disc 
(e.g. Wiseman et al. 2001; Lee et al. 2014, 2017b; 
Codella et al. 2014, Podio et al. 2015, Leurini et al. 2016), 
that is, blue-shifted towards 
the north-west and red-shifted towards the south-east.
Additional information is obtained from the channel maps 
(shown Fig. 4 with a spectral resolution of 1 km s$^{-1}$). 
The sizes and centroids of the emitting regions 
were obtained from elliptical Gaussian fits
in the $uv$ domain\footnote{We used the GILDAS $uv-fit$ task, which
allows fitting models directly through the $uv$ visibilities.
The error on centroid position provided by $uv-fit$ is the function
of the channel S/N and atmospheric seeing,
and is typically much smaller than the beam size.
As reported by Mart\'{\i}--Vidal et al. (2014), for instance,
an unlimited over-resolution
power can in principle be achieved if the dynamic range 
of the observations is arbitrarily large.}. 
The sizes (derived at $v_{\rm sys}$) are consistent
within the errors:

\begin{itemize}
\item
178(10)$\times$118(10) mas (72$\times$48 au; HDO); 
\item
166(14)$\times$100(16) mas (67$\times$41 au; CH$_3$CHO-A); 
\item
169(15)$\times$146(15) mas (68$\times$59 au; CH$_3$CHO-E).
\end{itemize}

At high velocities, we clearly spatially resolve
the red- and blue-shifted regions  
that move away from the jet axis. This reveales the velocity gradient. 
These spatial sizes overlap well with the disc 
traced by Lee et al. (2017a) using ALMA continuum
emission in Band 7 with a spatial resolution of 8 au (see Fig. 5). 
Although comparisons between images taken at different angular
resolutions have to be made with caution, 
Fig. 5 suggests that HDO and CH$_3$CHO
originate from a radius consistent with that of the dusty disc
around the protostar (60 au).

Figure 6 reports for HDO and CH$_3$CHO A\&E the emission 
centroid positions in each
channel, also obtained from elliptical Gaussian fits
in the $uv$ domain with the original 0.42 km s$^{-1}$
in order to better trace kinematics. 
The centroid distributions do not show 
a trend
expected for inner free-falling envelopes ($v$ $\sim$ r$^{-1}$) or 
discs ($v$ $\sim$ r$^{-1/2}$), where
at higher velocities the positions of the
blue- and red-shifted clumps approach each other.
We observe exactly the opposite trend (see Sect. 6
for more details on this aspect).
Conversely, this trend was indeed observed using
C$^{17}$O(3--2) 
and HCO$^+$(4--3), as observed at 0$\farcs$6 angular
resolution by Codella
et al. (2014) and Lee et al. (2014), respectively, 
revealing a rotating molecular disc/inner envelope.
Figure 5 further compares the spatial distribution of HDO 
and CH$_3$CHO at
velocities\footnote{
These velocities allow us to trace the disc
because at velocities closer to $v_{\rm sys}$, the emission
is dominated by the molecular static cloud.}
$v_{\rm sys}$ $\pm$ 2--3 km s$^{-1}$ with that of 
C$^{17}$O(3--2) as observed in the same dataset as presented here.
HDO and CH$_3$CHO are more compact than C$^{17}$O, which
is more elongated along the equatorial plane. This confirms
that these molecules trace different gas components (see Sect. 6). 

\begin{figure}
\centerline
{\includegraphics[angle=0,width=8cm]{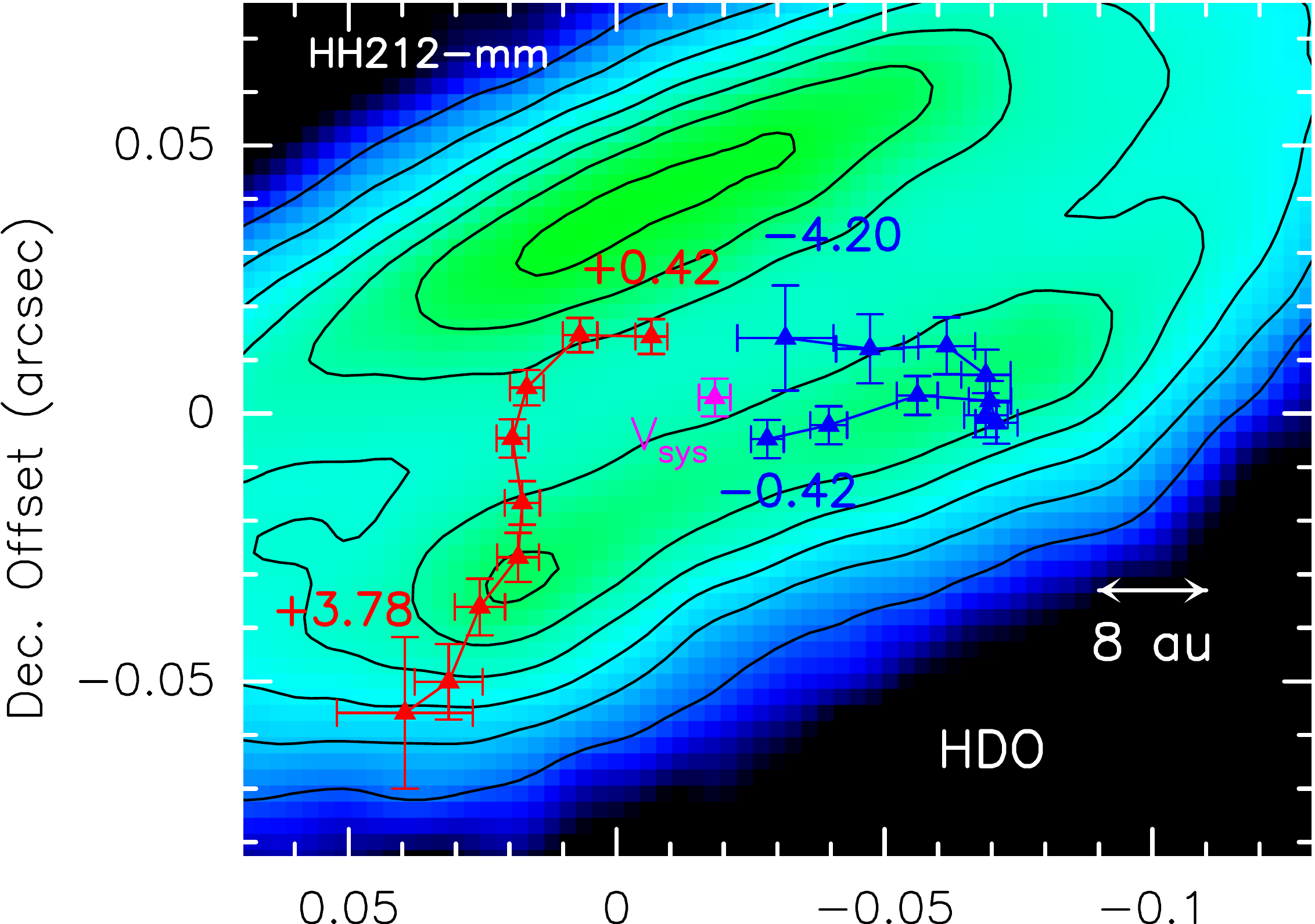}}
\centerline
{\includegraphics[angle=0,width=8cm]{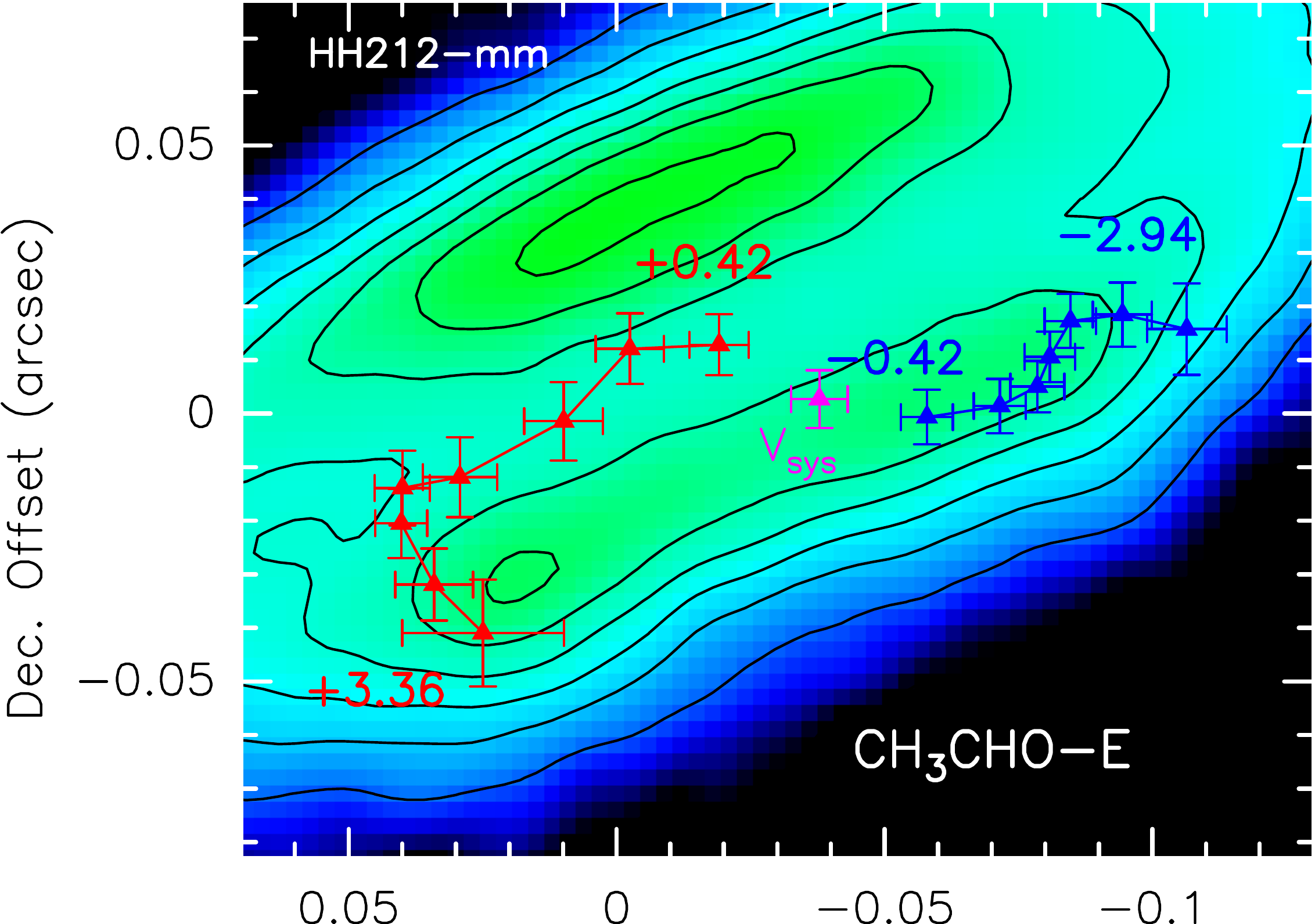}}
\centerline
{\includegraphics[angle=0,width=8cm]{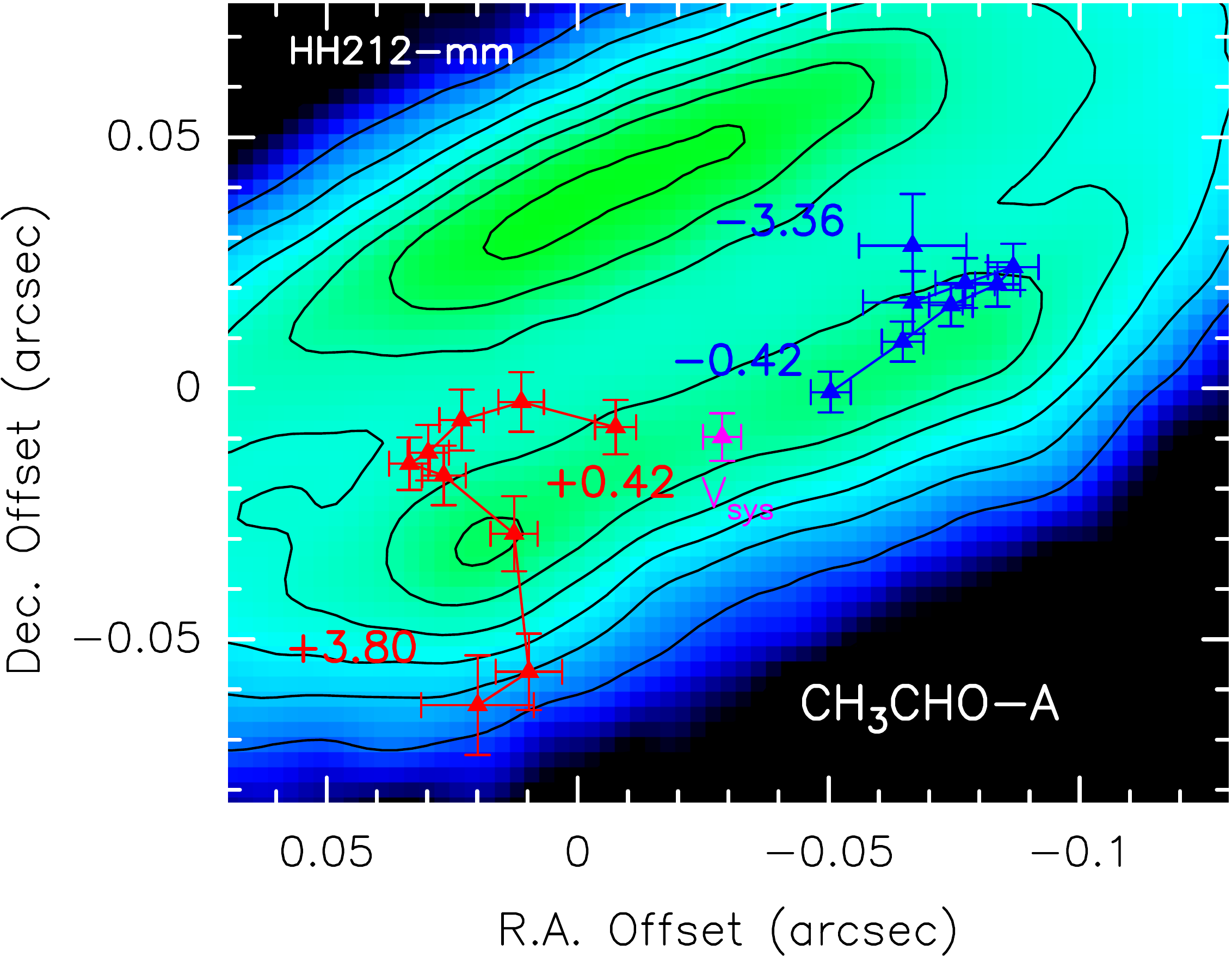}}
\caption{Distribution of the HDO(3$_{\rm 3,1}$--4$_{2,2}$) ({\it upper}), and
CH$_{\rm 3}$CHO(18$_{\rm 0,18}$--17$_{\rm 0,17}$) E\&A 
({\it middle and lower}) centroid positions 
(from fits in the $uv$ domain with 1$\sigma$ error bars) in the 
velocity channels (sampled at the original spectral 
resolution of 0.42 km s$^{-1}$).
Magenta points show the emission in the channel sampling
the systemic velocity, and 
red and blue datapoints denote the channels that are red- and blue-shifted in the velocity range (with respect to $v_{\rm sys}$), as identified by the labels.
The points are overlaid
on the disc traced by Lee et al. (2017a) using
ALMA Band 7 continuum observations (colour scale and contours).}
\label{uvfits}
\end{figure}

\section{Discussion}

We wish to answer the question about the origin of the hot, dense, chemically enriched,
and rotating gas around HH 212 mm. 
Our findings indicate 
that HDO and CH$_3$CHO emissions are related to 
the disc. The disc 
dominates the central mass (as seen in the continuum), and its sense of
rotation matches the velocity pattern seen in the observed lines.
Figure 7
reports the position-velocity (PV) diagrams 
of HDO and CH$_3$CHO along the equatorial plane (left panels) and the
jet axis (right). The velocity gradient associated with the disc
is clearly observed (quite similar to what was found 
using CH$_3$OH and CH$_2$DOH by
Lee et al. 2017b), while the PVs along the axis of the
jet do not show a conclusive trend.

\subsection{Origin of the iCOM emission}

It is instructive to compare our images with those reported
by Lee et al. (2017b):
the authors found that CH$_3$OH and NH$_2$CHO (among other species)
trace the surface layers (i.e. the atmosphere)
of the disc within the centrifugal barrier
located at $\sim$ 44 au.
We note that Bianchi et al. (2017), using the same ALMA dataset as we do here, found
very similar images for $^{13}$CH$_3$OH.
The iCOM emission detected by Lee et al. (2017b) 
probes two thin ($\sim$ 50 mas) layers above and
below the opaque mid-plane of the disc.
The maps reported here follow the iCOM vertical 
distribution up to about $\pm$ 100 mas
from the equatorial plane. A plausible explanation is that 
HDO and CH$_3$CHO trace the same surface layers as were detected
by Lee et al. (2017b), but 
convolved with a larger beam, which 
means that we see a single structure instead of two
thin layers. 
Once again, the comparison reported in Fig. 6 has to be
interpreted with caution, given the different angular resolution
of the two datasets. However, 
Figure 6 shows that the centroids (i.e. the barycenter
of the emitting size as measured in the $uv$ domain)
of the emission preferentially
peak towards the southern surface layers, in agreement
with Lee et al. (2017b), who found brighter methanol
and formaldeyde emission at this location. 
However, we cannot exclude that HDO and CH$_3$CHO
complement the CH$_3$OH and NH$_2$CHO images on
spatial scales larger than those used by Lee et al. (2017b),
hence tracing more extended emission.

The PV diagrams obtained from the HDO and CH$_3$CHO images
along the equatorial plane (Fig. 7, left panels)
show a linear velocity gradient that strongly suggests a narrow
rotating ring seen edge-on. 
In this case, the projected velocity is expected to be proportional
to the projected position offset from the protostar,
as is observed.
Other interpretations in terms of a strongly
self-gravitating structure with $M$ $\propto$ $r^{3}$ are
unlikely, as the dynamics of the infalling envelope
and of the inner disc in HH212 are better reproduced by a dominating
central mass $\sim$ 0.2 $M_{\rm \odot}$ (Lee et al. 2017c, 2018).
Figure 7 further indicates a radius for the emitting ring of
$\simeq$ 0$\farcs$15 (60 au), in agreement with the radius of
the centrifugal barrier of 0$\farcs$11--0$\farcs$12 derived at
higher resolution from HCO$^+$ infall kinematics by 
Lee et al. (2017b).
Hence, our results are consistent with a strong chemical enrichment 
in iCOMs in the ring where the infalling envelope
on the rotating disc, possibly due to low-velocity accretion shocks.
A similar effect has been observed in other more
evolved protostars
(see e.g. Sakai et al. 2014ab, 2016, 2017; Oya et al. 2016,
and references therein), which also supports this interpretation. Figure 5 shows that C$^{17}$O is
radially more extended than HDO and iCOMs: 
this supports that
the C$^{17}$O as observed at low velocities 
by Codella et al. (2014) using Cycle 0
images with a resolution of $\sim$ 0$\farcs$6 might 
trace the
inner portion of the infalling envelope
(the C$^{17}$O rotation curve is also consistent
with angular momentum conservation down to $\sim$ 0$\farcs$1
of the source, see the magenta curve
in their Fig. 3), while iCOMs (see Lee et al. 2017b and our
images) trace the outer region
of the molecular disc, which has then a radius of about 40--45 au.
On the other hand, because the iCOMs around the HH 212 protostar
are expected to be associated with low-velocity accretion shocks, they
are either formed directly in the external disc layers through gas-phase chemical reactions
or are directly injected into the gas-phase by thermal desorption
from grain mantles of an inner portion of the disc
(e.g. Lee et al. 2017b, Codella et al. 2017, and references therein, for a recent discussion).

\subsection{Low-velocity emission: outflowing and expanding rings}

Figures 4 and 5 show iCOM emission up to 40--50 au above and below
the disc plane, suggesting a vertical structure for the centrifugal 
barrier (see also Sakai et al. 2017). 
Are these iCOMs ouflowing, thus leaving the protostellar system?
Interestingly, all the three Cycle 4 HDO and CH$_3$OH images show 
emission peaks at low blue- and red-shifted
velocities, $\pm$ 1 km s$^{-1}$ from $v_{\rm sys}$, which 
look offset from the equatorial plane.
Although this result has to be verified using datasets with higher spatial resolution, it is interesting to note that this offset 
has previously been suggested 
by lower ($\sim$ 0$\farcs$6) angular resolution CH$_3$OH images
by Leurini et al. (2016) and was found in the $^{13}$CH$_3$OH maps
reported by Bianchi et al. (2017) using our Cycle 4 dataset. 
More specifically, when we focus on the low-velocity emission
of Fig. 6, 
the red-shifted emission lies to the north-east and the blue-shifted emission is
mainly detected to the south-west.
The observed offset is clearly inconsistent with the direction
of the jet (the blue-shifted lobe lies to the north-east and the
red-shifted lobe to the south-west, see Fig. 1). 
In addition, infall motions can clearly be ruled out
since we are observing the system so close to edge-on: 
any $v_{\rm z}$ infall motion 
is going to be projected at values too low ($\leq$ $\pm$ 0.2 km s$^{-1}$
at a radius of 40 au) to produce a shift $\pm$ 1 km s$^{-1}$ between
the red and blue peaks.
In such an edge-on view and at the current spectral resolution, 
we are mostly sensitive to the velocity component in the equatorial plane. 
In this case, the fact that maps at +1 km s$^{-1}$ peak in the 
northern (blue-shifted) lobe and the maps at --1 km s$^{-1}$ peak 
in the southern red-shifted lobe could be explained by {\it \textup{outflowing and expanding}} 
material that is observed against a strong optically thick disc (Lee et al. 2017c). 
The line excitation temperature ($\geq$ 80 K) is close to the dust temperature
(80--90 K; Lee et al. 2017c), therefore the line emission from the near 
side will not be detected on top of the continuum,
because it is negligible with respect to the background flux.
Figure 8 reports a sketch (not to scale) of the expected kinematics
in the two cases of outflowing and infaling material.  
Our images indicate that we observe only some portions of the 
outflowing and expanding gas, that is, 
{\it \textup{the far side, (red-shifted) in the upper part, and the near side 
(blue-shifted) in the lower part}.} 
If, in addition, the ring is rotating, the rotation motion will distort 
the gradient so that red-shifted emission peaks to the north-east and blue-shifted emission
to the south-west, as is indeed observed.

Interestingly, the trend observed in HDO and CH$_3$CHO
(and CH$_3$OH) at $\pm$ 1 km s$^{-1}$ is not observed in 
SO(10$_{\rm 11}$--10$_{\rm 10}$) and 
C$^{17}$O Cycle 0 data at the same velocities (see Fig.
5 of Podio et al. 2015 and Fig. A.1 of Codella et al. 2014).
A reasonable explanation is that SO and C$^{17}$O 
(Codella et al. 2014; Podio et al. 2015; Tabone et al. 2017)
probes larger regions than iCOMs, where  
optical depth effects against the dust disc are definitely diluted.

Finally, the spatial extent of the iCOMs structure is consistent
with the results
obtained for the L1527 protostar by Sakai et al. (2017):
the authors suggest that a fraction of the gas slowly moves away
from the disc mid-plane at the centrifugal barrier towards vertical directions.
Figure 7 indicates that this effect is more important for HDO,
which shows a
slightly wider distribution ($\pm$ 0$\farcs$2) than that of CH$_3$CHO.
The possibility
that HDO might be associated with outflowing motion has previously been proposed by Codella et al. (2016) based on the  
observed line profiles as obtained from ALMA 
data at lower ($\sim$ 0$\farcs$6) spatial resolution.

\begin{figure}
\centerline{\includegraphics[angle=0,width=8.5cm]{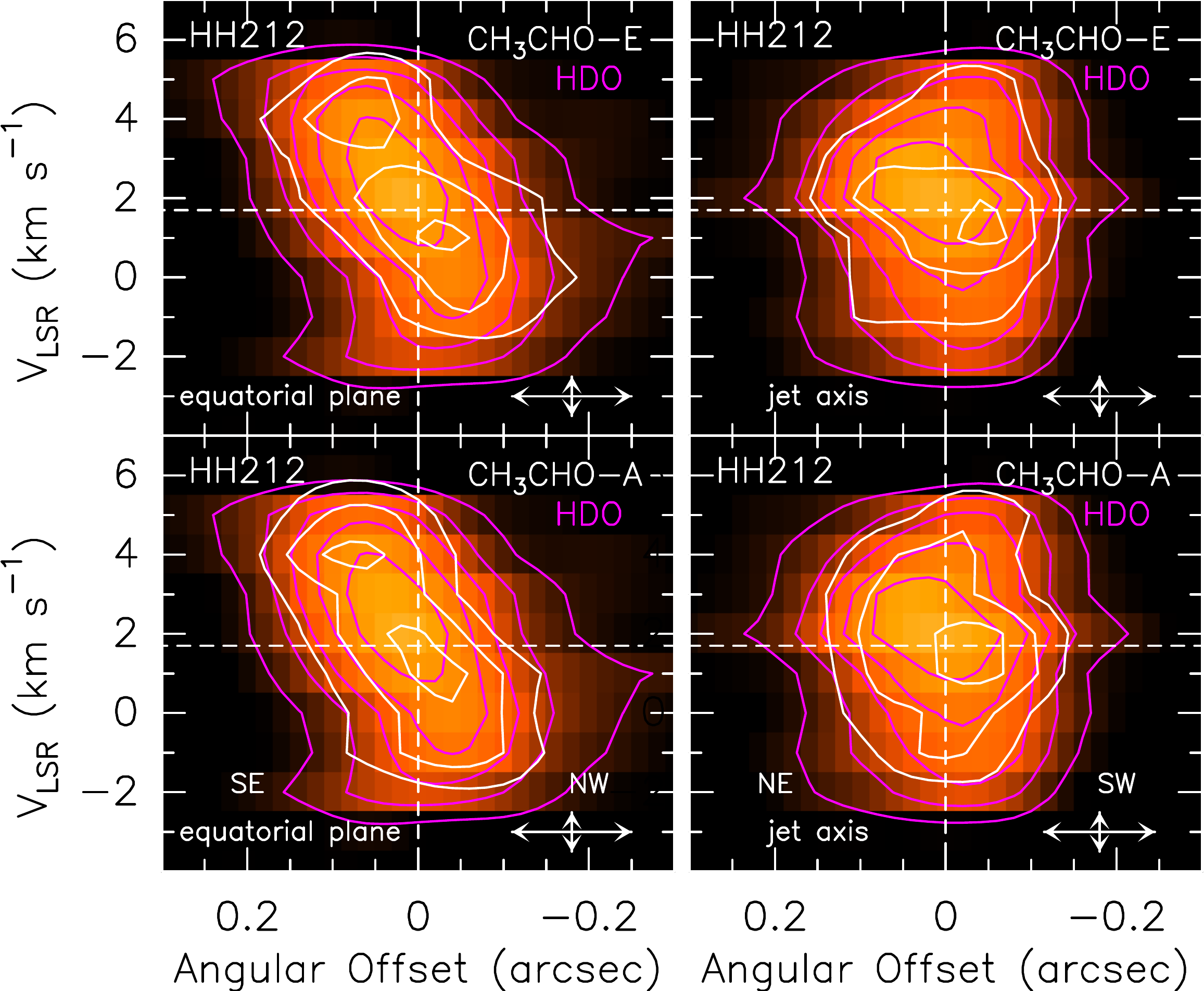}}
\caption{Position-velocity cut of HDO(3$_{\rm 3,1}$--4$_{2,2}$)
(colour scale, magenta contours) along (left panels) and perpendicular 
(right panels) to 
the equatorial plane
(PA = 112$\degr$, see Fig. 1) overlaid with the
CH$_{\rm 3}$CHO(18$_{\rm 0,18}$--17$_{\rm 0,17}$)-E
(upper panels) and CH$_{\rm 3}$CHO(18$_{\rm 0,18}$--17$_{\rm 0,17}$)-A
(lower panels), drawn as white contours.
The first contours and steps correspond to 3$\sigma$
(3.1 K km s$^{-1}$).
Dashed lines mark the position of the HH 212 mm protostar and the cloud
$V_{\rm LSR}$ (+1.7 km s$^{-1}$; Lee et al. 2014).
The error bars are drawn in the bottom right corners.}
\label{pv}
\end{figure}

\begin{figure}
\centerline{\includegraphics[angle=0,width=8cm]{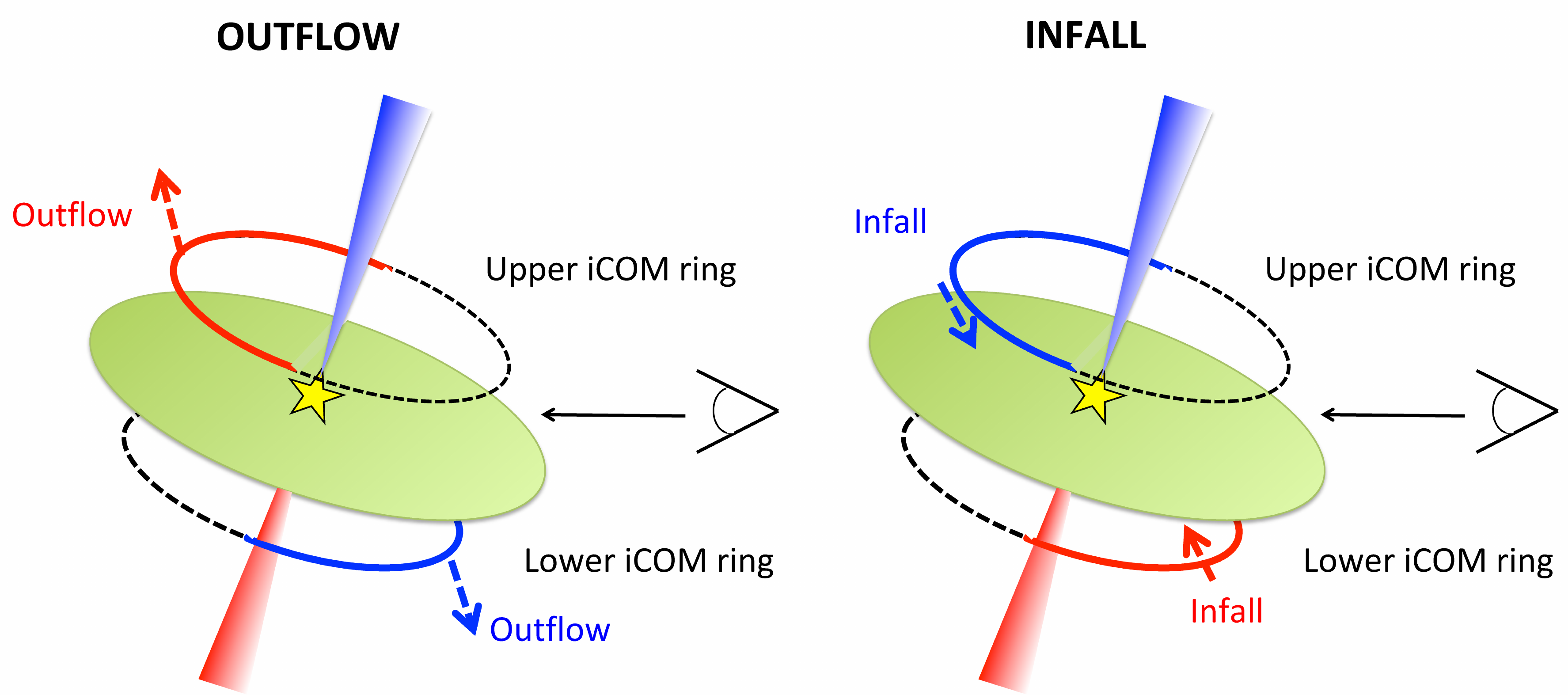}}
\caption{Cartoon (not to scale) illustrating the scenario expected 
in the inner 60 au region around the protostar for
low-velocity emission ($\leq$ 1 km s$^{-1}$ with respect to
$v_{\rm sys}$) in case
of outflowing (left) or infalling (right) chemically enriched gas (see text).}.
\label{cartoon}
\end{figure}

\subsection{Chemically enriched disc wind?}

Finally, it is intriguing to discuss whether 
our maps show signatures of
the disc wind, as suggested by Leurini et al. (2016), using
CH$_3$OH, Lee et al. (2017b), 
HCO$^+$, and by Tabone et al. (2017) and Lee et al. (2018),
SO and SO$_2$.
More specifically, SO and SO$_2$ appear to trace the outflow motion
farther away from the disc
(e.g. Fig. 3 of Tabone et al. 2017; see also Lee et al. 2018).
Hence the emission is collimated and fast, with 
the projected velocity $v_{\rm r}$ $\ll$ $v_{\rm z}$. 
On the other hand, by tracing HDO and CH$_3$CHO,
we could look closely at the plane of the disc.
The outflowing ring of gas might sample a flow 
that expands and 
is slower. 
Detailed models of magnetohydrodynamics disc winds are required to further analyse this aspect,
that is to say, to verify whether a portion of the HDO and CH$_3$CHO 
gas
traces a complementary part of the SO and SO$_2$
outflow. 

\section{Summary}

We presented Cycle 1 and Cycle 4 ALMA Band 7 observations
of HDO and CH$_3$CHO towards the HH 212 mm Class 0 object in Orion B.
The results confirm that
astrochemistry is a powerful tool for
imaging different kinematical components around protostars 
on solar system scales. 
The main findings are summarised below.

\begin{enumerate}

\item
Many (14) CH$_3$CHO emission lines (with
$E_{\rm u}$ in the 163--215 K range)  and 1 HDO line 
with $E_{\rm u}$ = 335 K have been imaged:
the emitting regions extend vertically over  
$\sim$ 0$\farcs$18 (73 au). 
All the detected profiles peak around the systemic velocity
and are $\sim$ 5--7 km s$^{-1}$ wide.

\item
The LTE analysis of the CH$_3$CHO emission leads to
a temperature of 78$\pm$14 K and a column density 
of 7.6$\pm$3.2 $\times$ 10$^{15}$ cm$^{-2}$. 
When we assume $N_{\rm H_2}$ = 10$^{24}$ cm$^{-2}$, the CH$_3$CHO abundance is 
$X_{\rm CH_3CHO}$ $\simeq$ 8 $\times$ 10$^{-9}$.
On the other hand, the LVG analysis of the HDO emission
at 335 K severely constrains the volume density,
n$_{\rm H_2}$ $\geq$ 10$^8$ cm$^{-3}$, and indicates 
column densities $N$(HDO) $\leq$ 3 $\times$ 10$^{17}$ cm$^{-2}$.

\item
CH$_3$CHO and HDO both show a velocity
gradient of about $\pm$ 2 km s$^{-1}$ with respect
to the systemic velocity along the equatorial plane.
The blue-shifted emission is located towards
the north-west, and the red-shifted emission lies towards the south-east,
in agreement with what has previously been found using other
molecular tracers of the inner protostellar region (such as 
C$^{17}$O, SO, HCO$^+$, CH$_3$OH, and NH$_2$CHO). 

\item
The CH$_3$CHO and HDO PV diagrams along the equatorial plane
show increasing velocities at increasing distance from the protostar. 
This is most likely the signature of emission from 
an edge-on ring of chemically 
enriched warm gas.
The radius is $\simeq$ 0$\farcs$15 (60 au), well in agreement with (i)
previous measurements based on other iCOMs such as NH$_2$CHO (Lee et al. 2017b),
(ii) the radius of the optically thick disc observed through
the continuum (Lee et al. 2017a), and (iii) the location
of the centrifugal barrier inferred from HCO$^+$ infall 
kinematics (Lee et al. 2017b). Hence, our data appear to probe
chemically enriched gas
where the infalling envelope meets the rotating disc,
where it probably creates low-velocity shocks.
Accurate modelling to investigate the chemistry
induced by low-velocity shock is needed.

\item
Our images show CH$_3$CHO and HDO emission 
at low velocities ($\pm$ 1 km s$^{-1}$ from $v_{\rm sys}$) up 
to 40--50 au above and below the disc plane.
This supports a vertical structure for the
centrifugal barrier, as has recently been found for the more evolved
L1527 protostar (Sakai et al. 2017). 
The observed spatial distributions 
could be the signature of two outflowing, expanding,
and rotating rings above or below the optically thick
equatorial disc plane, which could be a signature of
the basis of a disc wind. 
Clearly, further ALMA observations are needed to verify this possibility.

\end{enumerate}

\begin{acknowledgements}
We thank the referee, P. Ho, for valuable comments and suggestions.
This paper makes use of the ADS/JAO.ALMA\#2016.1.01475.S data (PI: C. Codella). ALMA is a partnership of
ESO (representing its member states), NSF (USA) and NINS (Japan), together with NRC (Canada) and NSC
and ASIAA (Taiwan), in cooperation with the Republic of Chile. The Joint ALMA Observatory is operated by ESO, AUI/NRAO and NAOJ.
We are warmly grateful to S. Yamamoto, N. Sakai, and Y. Oya for
instructive discussion on astrochemistry as applied to the HH212 system. 
We also thank M. Kounkel and L. Hartmann for valuable comments on
distances of the Orion cloud, and in particular of HH 212.
This work was supported
by (i) the program PRIN-MIUR 2015 STARS in the CAOS - Simulation Tools
for Astrochemical Reactivity and Spectroscopy in the Cyberinfrastructure
for Astrochemical Organic Species (2015F59J3R, MIUR Ministero
dell'Istruzione, dell'Universit\`a della Ricerca e della
Scuola Normale Superiore), (ii) the PRIN-INAF 2016  
"The Cradle of Life - GENESIS-SKA (General Conditions in Early Planetary Systems
for the rise of life with SKA)" and
(iii) the European Research Council (ERC) under the European Union's Horizon 2020 research and innovation programme, for the Project “The Dawn of Organic Chemistry” (DOC), grant agreement No 741002. 
MT and RB acknowledge partial support
from MINECO project AYA2016-79006-P. 
\end{acknowledgements}

\end{document}